# SDU

University of Southern Denmark

Standard energy data competition procedure

A comprehensive review with a case study of the ADRENALIN load disaggregation competition

Tolnai, Balázs András; Ma, Zheng Grace; Jørgensen, Bo Nørregaard

Go to publication entry in University of Southern Denmark's Research Portal

# Standard energy data competition procedure: A comprehensive review with a case study of the ADRENALIN load disaggregation competition

Balázs András Tolnai[1][0009-0004-4183-4340] and Zheng Ma[2][0000-0002-9134-1032] and Bo Nørregaard Jørgensen[3][0000-0001-5678-6602]

[123] SDU Center for Energy Informatics, Maersk Mc-Kinney Moeller Institute, The Faculty of Engineering, University of Southern Denmark, Odense 5230, Denmark
`bat@mmmi.sdu.dk`

**Abstract.** Crowdsourcing data science competitions has become popular as a cost-effective alternative to solving complex energy-related challenges. However, comprehensive reviews on hosting processes remain scarce. Therefore, this paper undertakes a detailed review of 33 existing data competitions and 12 hosting platforms, complemented by an in-depth case study of the ADRENALIN load disaggregation competition. The review identifies essential elements of competition procedure, including platform selection, timeline, datasets, and submission and evaluation mechanisms. Based on proposed 16 evaluation criteria, the similarities and differences between data competition hosting platforms can be categorized into platform scoring and popularity, platform features, community engagement, open-source platforms, region-specific platforms, platform-specific purposes, and multi-purpose platforms. The case study underscores strategic planning's critical role, particularly platform selection. The case study also shows the importance of defining competition scope which influences the whole competition content and procedure, especially the datasets.

**Keywords:** Data science competition, Data competitions, Competition Platforms, Competition timelines

## 1 Introduction

The phenomenon of crowdsourcing solutions to data problems through competitive platforms has grown immensely popular over the last decade. These competitions serve as a cost-effective alternative to traditional hiring, fostering a broad spectrum of innovative solutions by harnessing the collective intelligence of global participants. For competitors, these events offer a remarkable opportunity to learn new techniques, refine their skills, and augment their professional portfolios.

This trend is particularly vital in the energy sector, where data-driven solutions are crucial for addressing complex energy-related challenges and problems {Christensen, 2019 #62}. From energy efficiency and load forecasting to renewable energy integration and grid stability, data competitions play a significant role in generating groundbreaking solutions and accelerating the energy transition {Vanting, 2021 #109}.

Numerous companies and platforms, such as Kaggle, specialize in hosting these competitions. The selection ranges from free platforms to premium services, where professional teams aid in the competition's management. Choosing the hosting platform is one of the many pivotal decisions that underpin the successful execution of a data science competition.

While guidelines for hosting or setting up these competitions do exist, such as Kaggle's community competition setup guide [7] and Chalearn's guide [9], there is a notable lack of a comprehensive review that considers the entire process. This paper aims to fill this gap by thoroughly examining the hosting process of data science competitions, drawing insights from 33 prior competitions hosted in 2021 and 2022 by the NeurIPS 2022 conference (25 out of 33 competitions) [12] and other conferences on the AIcrowd platform. These competitions were collected in 2022 and they tackle a large variety of topics, including reinforcement learning, computer vision and forecasting.

To enrich our exploration, we present a case study - the ADRENALIN load disaggregation competition [14]. This competition is an integral part of the ADRENALIN project, a strategic initiative to crowdsource energy solutions for buildings. This paper provides a comprehensive review of standard energy data competition procedures and their importance in addressing energy-related challenges, with a spotlight on the ADRENALIN case.

The paper is organized as follows: The paper will first review the process of competition hosting. First, it looks at the official websites and platforms for competition hosting, and then analyses the stages, and durations of the competitions. Afterward, it looks into the technical parts of a competition, by reviewing the datasets, the starter kit, the submission, the evaluation, and the competition description. After the review, the paper showcases the case study of the ADRENALIN load disaggregation competition.

## 2 Review of Data Competition Procedures

The review of the 33 data competitions shows that a data competition includes the following 7 elements:

### 2.1 Official website

Every online competition necessitates an official website. These websites serve as the primary source for sharing up-to-date information about the competition, equipping participants with everything they need to compete. They provide background information, the problem statement, evaluation procedures, prizes, and details about sponsors and hosting organizations. They also offer guidance on the submission process, where to submit, and access to the dataset.

In addition to these websites, many competitions maintain a GitLab or GitHub page, as observed in 17 of the investigated competitions [17]. These platforms primarily function as repositories for sharing information. The competition description is usually incorporated in the readme.md file, displayed on the GitHub page. They can also host datasets, starting kits, reinforcement learning environments, or any other necessary tools. These resources can be effortlessly downloaded or forked by Git users. AIcrowd, for example, frequently utilizes its own GitLab to manage code submissions [17].

Among the 33 examined competitions, 27 maintained more than one website. However, it is not uncommon to solely use a data competition hosting platform for all purposes. Of the 33 competitions, six employed one of the hosting platforms as their official website. Of these, two used AIcrowd [18, 19], two used Kaggle [20, 21], and one-one used Codalab [22] and EvalAI [23].

## 2.2 Hosting platforms

Figure 1 shows the distribution of the used platforms in the 33 analyzed competitions. The figure shows that 10 of the competitions did not use a hosting platform. These used their own websites to organize the competition, handling submissions by uploading on the website, or through other means, such as Google Drive. Three of these were hosted by the Institute of Advanced Research in Artificial Intelligence [24-26], which has hosted multiple competitions on its own website. The most used platform was Codalab, used 8 times, followed by AI crowd, used 7 times. The other three platforms used in the sample of 33 competitions, were EvalAI 4 times, Kaggle 3 times, and DrivenData once.

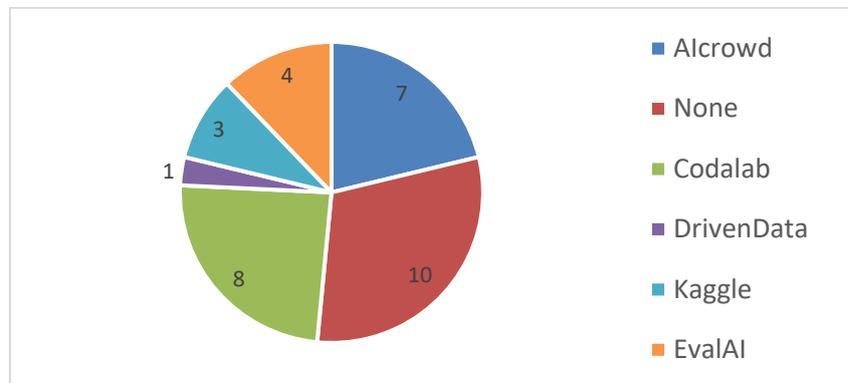

**Fig. 1.** Platforms used for the 33 reviewed data competitions.

There are many platforms designed to host data science competitions. One of the most important features of these platforms is the automatic evaluation of the submissions. Using the automatic scoring, a live leaderboard can be maintained throughout the competition. This helps competitors see how well their solution fares compared to the others. Since the evaluation is done automatically, it also ensures that the rankings are unbiased.

This paper applies 16 criteria (as shown in Table 1) to evaluate the 12 data competition hosting platforms, extending, and updating a comparison previously made by David Rousseau and Andrey Ustyuzhanin [27]. These hosting platforms are AiCrowd, CodaLab, CrowdAnalytiX, EvalAI, Kaggle, RAMP, Tianchi, Driven Data, Zindi, Topcoder, Bitgrit, and HackerEarth. This comparison is shown in Table 2.

**Table 1.** Platforms comparison criteria

| | Criteria | Description |
|---|---|---|
| 1 | Code-sharing | Code sharing gives participants an opportunity to share their work. This can help the community to find better solutions and increase the reproducibility of the winning solutions. |
| 2 | Code submission | Platforms that allow code submission can automatically run the submitted code to produce the predictions. Code submission has some advantages over result submission. |
| 3 | Active community | Does the platform have an active community, that will join the competition, and actively participate in it. |
| 4 | Staged challenge | Is it possible to create a competition with multiple different stages. |
| 5 | Custom metrics | Does the platform allow the hosts to define their own evaluation metrics or is it only possible to select from a list of predefined metrics. |
| 6 | Private evaluation | Private evaluation allows the competition hosts to evaluate the submissions without sharing the test data with the hosting platform. This is useful if the privacy of the test set needs to be protected. |
| 7 | Multi score leader board | Multi-score leaderboards allow the competitions to have multiple scores calculated from the submissions and posted on the leaderboard. |
| 8 | Human evaluation | Some sites allow the hosting of competitions, where automatic scoring is replaced with human-in-the-loop evaluation. |
| 9 | Open-source | Open-source platforms have higher transparency of their workings, and hosts have the option to set up their own servers by using the source code. |
| 10 | RL-friendly | Reinforcement learning is a unique type of ML, where the agents need to communicate with the environment. This requires a unique setup, not supported by all platforms. |
| 11 | Run for free | Some platforms allow the hosting of competitions free of charge. |
| 12 | Discussion forum | Most platforms provide a forum for each competition, where the contestants can communicate with each other and with the organizers. Common topics include sharing ideas, looking for teammates, or discussing issues they came across. |
| 13 | Technic support | Platforms marked as "certain" are designed for the users to set up technical support by themselves, relying only on the documentation. Despite this, the platform's creators and operators can be contacted directly with inquiries. |
| 14 | Arrangement and management services | Does the platform provide services for arranging and managing data competition events. |
| 15 | Ease of use for hosting | How easy is it to host and set up a data science competition on the platform. |

| 16 | Ease of use for participation | How easy is it to join and participate in a competition. This includes registration, submission, etc. |

The comparison shows that the similarities and differences among the reviewed 12 platforms can be categorized into:

**Platform Scoring and Popularity**. In this comparison, Kaggle was the highest-scoring platform, with 14 points. As Kaggle is the most known and most used platform, competitions hosted here normally attract the most participants. AIcrowd follows as the second scoring 13.5 points. With 13.25 points, the third competition in ranking is Eval.AI, one of the fully open-source and free competition hosting platforms on this list. Ranking fourth with 12.75 points is Codalab. Driven Data also has an active community, though not quite as large as some other platforms.

**Platform Features**. Kaggle has also become more flexible over time, and competitions of all categories can be found on the platform. It is important to note, that Kaggle has both free and paid competitions. A free competition is very easy to host, but it has some restrictions compared to the paid version and would score lower. This version lacks some features, such as custom metrics (although there are many metrics to choose from), human evaluation, and technical support. AIcrowd does not have a free version. It is a flexible platform, with many different competitions hosted, which is the main reason for its high score. Codalab allows hosts to make significant changes to almost all aspects of the competition after its initial setup, using the dashboard on the website.

**Community Engagement.** Kaggle, AIcrowd, and Driven Data have active communities, with Kaggle attracting the most participants. AIcrowd usually has a few hundred competitors, and a few threads in the discussion forums. Driven Data, while not as large as some other platforms, hosts large competitions with substantial prize money.

**Open-source Platforms**. Eval.AI and Codalab are fully open-source and free competition hosting platforms. Eval.AI offers an email where it is possible to get in contact for those who have questions or have run into problems with the setup. A documentation can also be found, which includes a guide on how to set up competitions. Codalab also has good documentation online, and multiple example competitions can be found on Codalab's GitHub page.

**Region-Specific Platforms**. Tianchi, operated by Alibaba Cloud, is considered the Chinese equivalent of Kaggle. Ramp is an open-source platform, developed by the Paris-Saclay Center for Data Science and is mainly used by the University of Paris-Saclay to host competitions. Zindi is a platform that aims to support the African data science community.

**Table 2.** Comparison of data competition event hosting platforms

| Criteria | AI-Crowd [1] | CodaLab Competitions [2] | CrowdAnalytiX [3] | Eva-lAI [4] | Kaggle [5] | RAMP [6] | Tianchi [8] | DrivenData [10] | Zindi [11] | Topcoder [13] | Bitgrit [15] | HackerEarth [16] |
|---|---|---|---|---|---|---|---|---|---|---|---|---|
| 1 | Y | Y | X | Y | Y | Y | Y | Y | Y | X | X | ? |
| 2 | Y | Y | X | Y | Y | Y | Y | Y | X | Y | X | Y |
| 3 | *** | *** | * | ** | **** | * | **** | *** | *** | ** | ** | ** |
| 4 | Y | Y | X | Y | Y | Y | Y | Y | X | X | Y | Y |
| 5 | Y | Y | Y | Y | Y | Y | ? | Y | ? | X | ? | ? |
| 6 | Y | Y | X | Y | X | X | X | Y | ? | X | ? | ? |
| 7 | Y | Y | X | Y | X | X | X | Y | X | X | X | X |
| 8 | Y | X | X | Y | X | Y | X | X | X | X | X | X |
| 9 | X | Y | X | Y | X | X | X | X | X | X | X | X |
| 10 | Y | Y | X | Y | Y | X | X | X | X | X | X | X |
| 11 | X | Y | X | Y | X | X | ? | X | X | X | X | X |
| 12 | Y | Y | X | Y | Y | X | Y | Y | Y | Y | Y | Y |
| 13 | Y | Y (certain) | Y | Y (certain) | Y | Y | Y | Y | Y | Y | Y | ? |
| 14 | Y | X | Y | X | Y | X | Y | Y | Y | Y | Y | Y |
| 15 | **** | ** | **** | * | **** | *** | **** | **** | **** | **** | **** | **** |
| 16 | *** | ** | ** | ** | **** | ** | ** | *** | *** | ** | ** | ** |
| Total score | 13.5 | 12.75 | 5.75 | 13.25 | 14 | 8.5 | 9.5 | 11.5 | 6.5 | 6 | 6 | 6 |

The last row is a score calculated from the other criteria. It was calculated as every Y is one point, and every * is 0.25 points.

**Platform-Specific Purposes**. Kaggle, Tianchi, and Zindi are not only about the competition but a central hub for learning and networking. The website for Tianchi serves as a sharing point for datasets, data science learning, and getting in touch with other computer scientists through the forums. Zindi also facilitates learning and job searching.

**Multi-purpose Platforms**. Topcoder, unlike the other platforms, is not specialized to host Data science competitions. It offers different tracks, such as ones called development, design, and QA (Quality assurance).

## 2.3    Competition descriptions

The information shared on the websites and the official competition platform is important, as the participants read these to decide if they want to join the competition, or not. This is also their source of information, for what the competition is about, what they should do, and what they should expect from the competition. Competition descriptions usually have the following parts:

**Background information**. This explains what the purpose of the competition is and what is the importance of the task the participants are trying to solve.

**Problem formulation** explains what the task is, and what is the target column in the dataset. It needs to be clear, so the participants know what they are going to be working on for the duration of the competition.

**The timeline of the competition** describes all of the deadlines, starts, and end dates of each competition phase.

**The prizes section** describes what prizes are available at the competition. This is usually prize money, split between the top finish contestants, but it can also be a publication opportunity or a conference or workshop invitation. In one competition the top finishers got offered a position, and in another one items were given out as special prizes.

**The dataset description** explains what the available dataset contains. Usually explains each file, and each column in the dataset.

**The submission section** describes the submission process, whether it is code or result, what needs to be uploaded, and where.

**A description of the evaluation process** usually discusses the evaluation metric, and how the private evaluation takes place.

Competitions usually need **a set of rules**, this usually asks the competitors to avoid things that could be considered cheating. Such as privately sharing code to take multiple top positions, and trying to find ways to abuse the submission system, such as creating multiple accounts to circumvent the submission limits.

## 2.4    Timeline and stages

Competitions have a timeline, marking the most important dates of the competitions. Although the timelines are normally set before the competition starts, deadline extensions do occasionally happen. Competitions can have multiple stages, with different data availability, and different submission types.

**Pre-phase.** Some competitions have a stage before the main event of the competition takes place. In this study, these are referred to as a "pre-phase". Out of the analyzed competitions, 13 had a pre-phase. Twelve of these were an announcement, or a similar pre-phase with a different name. These usually include the release of part of the data, such as the training set without the test set, the release of the starter kit, and the announcement of the full details of the competition. This includes phases with different names, but similar purposes, such as quick start, registration, data, or starting kit release.

Two competitions have a warm-up phase, which this study also considers as a pre-phase. During this phase, most necessary information is already released, but the leaderboard is either not open, or the submissions are tested against the training data.

These phases are usually intended to help competitors familiarize themselves with the dataset, and the problem at hand. It is also a good opportunity for the organizers, to take some feedback from the participants, and potentially make adjustments before the competition starts at its earnest.

**Main phases.** Eight out of the 33 competitions have only one main phase. These are the simplest competitions. For example, the majority of the competitions hosted on Kaggle fall into this category. These competitions usually have a start date, a final submission deadline, and often a team merger deadline or a rule acceptance deadline.

Seven of the analyzed competitions have 2 stages. In the second stage of the competition, the organizers often release additional data. This can serve multiple purposes, such as lowering the time participants get with the entire dataset and lowering the ability to purposefully overfit the dataset. Competitions often have a final phase, where the evaluation is held on a previously not used part of the dataset, with a very limited number of submissions. This is necessary to make sure that it is protected from leaderboard probing, and that the models are not overfitting by being finetuned, based on the results of the submission.

Some platforms, notably Kaggle, use the Private leaderboard system instead of a separate private evaluation round. In this setup, the test set is separated into a private and a public subset. Submissions are made on the entire test set, but feedback is only provided about the results on the public part. After the end of the competition, the leaderboard based on the private part is released, which reflects the results of the competition. The Ariel Machine Learning Data Challenge [28] competition had both a final submission and a second data release phase, making it the only three-phase competition with standard phases.

Competitions can have unique phases, and sometimes have multiple tracks. PETs Prize Challenge: Advancing Privacy-Preserving Federated Learning [29] had three phases unique to the problem. Participants could register as either a team that created a privacy-preserving federated system, or a team that tested these systems by trying to devise attacks on them. In the first phase, participants had to submit a concept paper for their federated system ideas. In the second phase, these concepts were developed and scored by judges, while the last phase was for the testing of these systems, by the tester teams. The Reconnaissance Blind Chess [30] was a competition, that was hosted as a tournament. There were two, optional test tournaments, and a real tournament. The Real

Robot Challenge 2022 had a simulated qualification phase first, which was designed to limit the number of runs necessary on the real robots, in the second phase.

Competitions have multiple tracks, when there are multiple, closely correlated problems the organizers would like to solve. In these cases, often they chose to organize one competition, with multiple tracks, instead of a series of separate ones. The NeurIPS 2022 IGLU Challenge [18] had two tracks, the first one was about creating an AI that can follow instructions given in a natural language, to build structures. In the second task, an AI had to ask clarifying questions, when the instructions given are not enough to construct the structure. The Trojan Detection Challenge [31] had 3 tracks for three tasks. The first task was to identify trojaned networks while the second task was to classify trojaned networks into different categories. The last task was to create trojaned networks, that are difficult to identify. The Neural MMO challenge [32] had two tracks, in one of them the agents had to play the game alone, while on the other track, the agents had to play the game together, interacting with each other.

Most competitions have some form of activities after the competition has ended. This includes the announcement of the winners and contacting the winning teams to hand in further information about their solutions, such as documentation, papers, and program code. Some competitions are followed by a conference, or a workshop, where the winning teams are invited to present their solutions.

## 2.5 Competition durations

The duration of each competition can vary, sometimes with large differences. The length of the one main stage competitions can be seen in Table 4. Their average length is about 77 days. Commonly the main stage of the competitions is around 90, roughly three months.

It is shown in Table 5. That two-stage competitions usually have a longer first stage. On average the length of the first stage is 74 days, while the second stage is only 19 days. This is especially true when the second stage is the private test phase. This usually does not need a long time, as it is intended for submission of final models previously created, rather than for further development. Overall, the length of the competitions tends to be somewhere between two and four months.

**Table 4.** Duration of One-stage competitions

| Event | Days |
|---|---|
| Discover the mysteries of the Maya - ECML PKDD 2021 - Discovery Challenge [33] | 91 |
| Feedback Prize – Predicting Effective Arguments. [20] | 91 |
| RSNA 2022 Cervical Spine Fracture Detection Identify cervical fractures from scans [21] | 91 |
| Global Challenge 2021 [34] | 61 |
| ADDI Alzheimer's Detection Challenge [35] | 43 |
| BASALT Competition 2022 [19] | 98 |
| OGB Large-Scale Challenge (OGB-LSC) [36] | 161 |
| AutoML Decathlon 2022 [37] | 118 |
| Natural Language for Optimization (NL4Opt) NeurIPS 2022 [38] | 92 |
| Second AmericasNLP Competition: Speech-to-Text Translation for Indigenous Languages of the Americas [39] | 10 |
| Open Catalyst Challenge [23] | 16 |
| Multimodal Single-Cell Integration Across Time, Individuals, and Batches [40] | 92 |
| Sensorium 2022 Competition [41] | 117 |
| Visual Domain Adaptation Challenge [42] | 10 |
| Habitat Rearrangement Challenge 2022 [43] | 59 |

**Table 5.** Duration of two-stage competitions

| Event | Stage 1 (days) | Stage 2 (days) |
|---|---|---|
| CityLearn Challenge 2022 - Multi-Agent Reinforcement Learning for energy management in cities [17] | 46 | 30 |
| EURO Meets NeurIPS 2022 Vehicle Routing Competition [44] | 91 | 27 |
| Weather4cast Multi-sensor Weather Forecast Competition [26] | 110 | 6 |
| Data Purchasing Challenge 2022 [45] | 24 | 35 |
| Cross-Domain MetaDL 2022 [46] | 62 | 30 |
| 2022 NeurIPS Driving SMARTS Competition [47] | 92 | 10 |
| Weather4cast 2022 [24] | 80 | 8 |
| MyoChallenge [48] | 69 | 6 |
| Weakly Supervised Cell Segmentation [49] | 93 | 19 |

## 2.6 Data and starter kit

**Starter kit.** Starter kits often accompany competitions, proving particularly crucial for those involving complex submission requirements, code submissions, and notably, reinforcement learning contests. These kits should contain all the necessary resources for making a submission, generally including a functional example. Occasionally, a baseline model may be included to provide competitors with a solid starting point, encouraging them to modify the baseline model according to their approach. The primary aim of these kits is to ensure that participants can correctly submit their entries by swapping the model in the starting kit with their own.

Out of the 33 competitions analyzed, 26 provided a starter kit. Sixteen of these were available through GitHub, four via GitLab, and the remaining six could be found on the hosting platform or website.

**Dataset.** Datasets are usually shared through the Competition hosting platforms, as they normally offer the possibility to upload the dataset on their platform, where the competitors can download it. Some competitions use other ways for dataset sharing, for example, GitHub or GitLab. The dataset is normally available alongside the starter kit or example submission.

The dataset is one of the most important parts of every data science problem, and as such, it is important for hosting a data science competition as well. The organizers must make sure that there is sufficient data available to teach machine learning models. A general rule of thumb is that the more data is the better. There are a few additional things that must be kept in mind when creating the dataset for the competition:

- In standard data science competitions featuring a private testing phase, the dataset must be divided into three parts: training, public validation, and private validation. When the competition employs a public and private leaderboard system, only two splits are necessary - one for training and one for validation. The training split should include the solutions column.
- In high-stakes competitions with monetary rewards, itis crucial to keep the test set concealed and devoid of leaks. If some participants gain access, it could compromise the competition's results.

## 2.7   Submission and evaluation

**Submission.** There are two main ways a competition can take submissions. These are result submission and code submission. Out of the 33 competitions looked at in this paper, 16 used result submission, and 14 used code submission. Additionally, The Real Robot Challenge 2022 [50] used a system, where in the first round, participants had to submit their achieved score, and the code was only rerun for the top teams to verify their score. The Reconnaissance Blind Chess competition [30] did not require submission, as the tournament was played online. If a competition uses result submission, the test data must be shared with the participants, without the target column. The competitors have to locally compute their predictions, and submit it in tabular format. Afterward, the submission can be directly evaluated against the ground truth.

With code submission, the participants must submit their code, which is re-run on the competition's server to compute the predictions. This requires more computational power from the server. In exchange, it allows the test set to remain entirely hidden. In addition, it also allows further options for the evaluation of the submission, notably the measurement of the code's runtime. Code submission competitions normally have a runtime limit. This has the technical purpose of not allowing single submission to occupy a computational unit for an indefinite duration, but it can also help create more practical models, for real-life applications.

**Evaluation.** Normally, evaluation is the automatic calculation of a score, by comparing the predictions with the ground truth. Usually, there are two test sets, to calculate a public, and a private score. The public score serves as feedback for the competitors, which appears on the public leaderboard. The private score is used to calculate the outcome of the competition. This is either computed parallelly with the public score, at each submission or calculated in a final, separate round. This is important to avoid overfitting on the test data. The public dataset is also vulnerable to leaderboard probing, where the participants use various methods to infer information about the data based on the returned scores. This can in extreme cases lead to revealing the ground truth.

The comparison of the ground truth and the prediction is done via a mathematical formula, often called metric, or score. There have been many metrics developed over the years, with different goals in mind. For example, a simple F1 score is used in [22], and mean intersection over union for evaluation is used in a computer vision competition [42]. Furthermore, in the CityLearn Challenge [17], the average of 6 metrics is calculated in order to evaluate electricity consumption and carbon emissions. These metrics include ramping, 1-load factor, average daily peak demand, maximum peak electricity demand, total electricity consumed and carbon emissions. These Metrics are not unique to competitions, as it is always desirable to measure the accuracy of the machine learning model, in a way that they can be compared to each other. Loss functions are similar, often the same formulas, which are used during the learning process of an algorithm, though they usually have different criteria to fulfill. As loss functions are used during the training process, they usually need to be fast to compute, and gradient descent-based learning can only be performed on differentiable formulas. Evaluation metrics usually aim to be easy to understand for humans, and to capture and emphasize the most important aspects of the model.

## 3 Case study

This case study centers around the organization of the "ADRENALIN 2023: Building Energy Load Disaggregation Challenge", a competition embedded in the ADRENALIN project with the aim of developing energy load disaggregation algorithms.

### 3.1 Competition scope definition

The overall theme of the competition is non-intrusive load disaggregation. Load disaggregation deals with the problem of discovering the sub-loads that make up a central, aggregated load. This can be beneficial for multiple reasons, e.g., to propagate energy saving behavior, or to create more precise demand response algorithms {Ma, 2021 #110}.

To determine the scope of the competition, a scoping review was also conducted [51]. Based on this review, it was decided that the competition will use low-frequency data, and aim to produce lightweight models. Furthermore, the literature also shows that the transferability of the developed forecasting models in buildings is a challenge, therefore the competitions also aim to test the applicability of the proposed solutions on different buildings.

### 3.2    Official website

Similarly to the majority of the analyzed competitions, the case will also have multiple websites. The official website of the competition (https://adrenalin.energy/adrenalin-2023-building-energy-load-disaggregation-challenge) is on the ADRENALIN project's website [14]. The competition's other website is hosted on Codalab, which will handle submissions and maintain the leaderboard. The description of the competition's details is available on both sites.

### 3.3    Hosting platforms

The selection of a hosting platform necessitated an evaluation of various options. Given the project's limited funding, the decision was made to utilize a free hosting service. Kaggle, known for its user-friendly interface, was one of the options, along with Eval.AI, and Codalab. However, Kaggle's restrictions, including the prohibition of prize offerings in the free version, led to its exclusion. After experimenting with Codalab and Eval.AI, Codalab was chosen due to its relative simplicity.

### 3.4    Competition timeline, stages, and durations

Based on prior investigations, a three-stage competition was designed, which comprises a pre-phase, and two main phases. It commences with a **warm-up phase** from the 1st of February to the 15th of March 2024. This gives about 1.5 months (43 days) for participants to familiarize themselves with the problem.

The second phase is the **development phase**, stretching from the 15th of March to the 14th of June 2024, lasting 91 days. This corresponds to the duration of the development phase in the reviewed competitions. This is when participants refine their models, experiment with solutions, and struggle to achieve the best solution.

Given that Codalab does not support a private leaderboard system, the competition ends with the **private test phase**, lasting from the 15th of June to the 31st of July. This 1.5 month gives plenty of time for the participants to make a final submission, and for the organizers to validate the submissions. During this phase, the submissions are evaluated against previously unseen data to test their ability to generalize. This phase determines the final leaderboard positions and competition winners.

### 3.5    Data and Starter kit

**Starter kit.** The competition will feature a starter kit. This will include an example of a correct submission, and the evaluation program that will be used on Codalab to calculate the scores. This will help contestants locally test their submissions, ensuring their correctness. The starter kit might also include a working baseline model, chosen based on a scoping review, as an example.

**Data.** The competition will feature a diverse dataset, collected by the ADRENALIN project partners, from various countries across the world. This will provide a sufficient amount of data to facilitate learning. At least three datasets will be created: training data, test data for the development phase, and test data for the private test phase.

The dataset will include main- and sub-meter measurements from different buildings, including multiple different features, using low-frequency sampling. Supplementary information on the building's properties and weather information will also be included where available.

Furthermore, the usage of external data is allowed in this competition, as long as it is free and publicly available. The competitors must ensure that all data they use is freely available to all participants, and post access to the dataset on the competition forum before the end of the competition.

### 3.6 Submission and evaluation

**Submission.** The submission of solutions will take place through the Codalab platform, using Codalab's submission system. The submission will be a result submission. Result submission is generally simpler to set up and guarantees that the submissions will have the same result as on the participant's own computer.

**Metric.** Deciding the metric for the competition is not a simple task. To see what evaluation metrics are used, a further study was conducted on a large number of competitions (130), to determine what are the most popular metrics used in current data science competitions. Furthermore, the literature review of load disaggregation also revealed the most popular metrics in the scope.

The most relevant metric was determined to be mean absolute error (MAE). MAE can be used to calculate the average difference between the predictions and the ground truth for the entire predicted time period. Based on the literature, it is preferred over Root mean squared error, as it does not emphasize larger errors disproportionately compared to smaller errors. Since Codalab can display multiple scores on the leaderboard, it is possible to add further metrics. Relative Mean Absolute Error and Mean Absolute Percentage Error are two averaged versions of MAE, which can provide further information, for human interpretation of the results.

### 3.7 Competition descriptions

The descriptions of the competition will be available through both the official website and the Codalab platform. The descriptions include the sections identified during the revision. It starts with the background, explaining the importance of energy disaggregation. This is followed by the problem statement, stating the goals and the task of the competition, based on the literature review. The following dataset description first gives an overview of the datasets, discusses its main features, and finally each file, and their contained features separately. The submission and the evaluation and metrics sections present the submission and the evaluation processes. The timeline section displays the information presented earlier. To help with getting in touch, a link to the discussion forum is also provided. Finally, the rules of the competition are provided, and acknowledgment to the sponsors of the competition.

## 4 Discussion and Conclusion

This paper conducts a comprehensive review on the procedure of data science competitions. This review uncovers the key elements involved, which include the construction of official websites, the selection of hosting platforms, timelines, datasets, starter kits, submission and evaluation protocols, and competition descriptions.

In a comparative analysis of various data science competition platforms, Kaggle was found to be the highest-scoring and most popular platform due to its flexibility, variety of competitions, and active community. AIcrowd, Eval.AI, Codalab, and Driven Data also scored high, offering a range of features and community engagement opportunities. Eval.AI and Codalab stand out as open-source platforms, while Tianchi, Ramp, and Zindi cater specifically to certain regions. Kaggle, Tianchi, and Zindi also serve as hubs for learning and networking, extending beyond just hosting competitions. Unlike the other platforms, Topcoder has a broader focus, offering tracks in development, design, and quality assurance alongside data science.

The case study of organizing a competition such as the Building Energy Load Disaggregation Challenge reveals the importance of careful planning and strategic decision-making in structuring the competition, choosing the hosting platform, and defining the submission and evaluation mechanisms. Especially the definition of competition scope which are not yet discussed in the literature. The competition scope influences the whole competition content and procedure, especially the datasets. The choice of Codalab for this competition exemplifies how factors such as cost and usability can significantly influence the hosting platform selection.

The insights gained from this study carry important implications for stakeholders in the energy sector seeking to organize data competitions to address pressing challenges. These stakeholders can range from academic institutions, industry professionals, government entities, to non-profit organizations. By applying the findings of this paper, they can more effectively set up competitions that not only crowd-source innovative solutions but also foster a culture of learning and creativity within the sustainable energy domain.

However, it is important to consider the limitations of this study. As it only samples 33 competitions, it captures just a fraction of the data competitions being hosted. Future research should include a larger sample size and a wider range of competitions to provide a more comprehensive understanding of best practices in data competition hosting.

## Acknowledgment

This paper is part of the Project "Data-dreven smarte bygninger: data sandkasse og konkurrence" (Journal-nummer: 64021-6025) funded by EUDP (Energy Technology

Development and Demonstration Program) and IEA EBC Annex 81 Data-Driven Smart Buildings project funded by EUDP (case number: 64019–0539), Denmark.